\documentclass[twocolumn]{aastex63}

\usepackage{amsmath}
\usepackage{verbatim}

\submitjournal{ApJ}

\graphicspath{{./}{figures/}}

\begin{document}

\title{Beyond runaway: initiation of the post-runaway greenhouse state on rocky exoplanets}

\correspondingauthor{Ryan Boukrouche}
\email{ryan.boukrouche@physics.ox.ac.uk}

\author[0000-0002-5728-5129]{Ryan Boukrouche}
\affiliation{Atmospheric, Oceanic and Planetary Physics, Department of Physics, University of Oxford, UK}

\author[0000-0002-3286-7683]{Tim Lichtenberg}
\affiliation{Atmospheric, Oceanic and Planetary Physics, Department of Physics, University of Oxford, UK}

\author[0000-0002-5887-1197]{Raymond T. Pierrehumbert}
\affiliation{Atmospheric, Oceanic and Planetary Physics, Department of Physics, University of Oxford, UK}

\begin{abstract}

% BACKGROUND (1 sentence)
The runaway greenhouse represents the ultimate climate catastrophe for rocky, Earth-like worlds: when the incoming stellar flux cannot be balanced by radiation to space, the oceans evaporate and exacerbate heating, turning the planet into a hot wasteland with a steam atmosphere overlying a possibly molten magma surface.
% OBJECTIVES+METHODS (1-2 sentences)
The equilibrium state beyond the runaway greenhouse instellation limit depends on the radiative properties of the atmosphere and its temperature structure. Here, we use 1-D radiative-convective models of steam atmospheres to explore the transition from the tropospheric radiation limit to the post-runaway climate state. To facilitate eventual simulations with 3-D global circulation models, a computationally efficient band-grey model is developed, which is capable of reproducing the key features of the more comprehensive calculations. 
% RESULTS (1 sentence)
We analyze two factors which determine the equilibrated surface temperature of post-runaway planets.  The infrared cooling of the planet is strongly enhanced by the penetration of the dry adiabat into the optically thin upper regions of the atmosphere. In addition, thermal emission of both shortwave and near-IR fluxes from the hot lower atmospheric layers, which can radiate through window regions of the spectrum, is quantified. 
% CONCLUSIONS (1 sentence)
Astronomical surveys of rocky exoplanets in the runaway greenhouse state may discriminate these features using multi-wavelength observations. 
\end{abstract}


\section{Introduction} \label{sec:intro}

The temperate orbital region where liquid water can be stable on the surfaces of rocky exoplanets, often referred to as the habitable zone, is bounded starward by the runaway greenhouse limit \citep{1993Icarus..101..108K}. It defines an orbit inside which a terrestrial planet harboring a secondary atmosphere and a reservoir of water can fall into a positive feedback loop and lose its ability to sustain habitable surface conditions \citep{2013ApJ...765..131K}. When the instellation exceeds levels at which the planetary climate can radiatively equilibrate, the available condensed water present at the surface evaporates into the atmosphere. This addition of water vapor to the increasingly steam-dominated atmosphere drives a positive feedback loop that warms the planet and accelerates the evaporation process until all available surface water is exhausted in the post-runaway stage. 

Planets undergoing a runaway greenhouse stage reach surface temperatures well in excess of those that Venus and Mercury experience in our Solar System today. Mercury, with its tenuous exosphere, reaches about $740 \: \mathrm{K}$ on its day side. Venus, with its dense atmosphere, maintains an average surface temperature of $737 \: \mathrm{K}$. Venus's orbit is within the runaway greenhouse instellation threshold, and one of the main evolutionary scenarios proposed is that it went through a moist and/or runaway greenhouse phase early in its history \citep{Kasting1987,Goldblatt2013,Leconte2013}, which would explain the absence of water and the massive $\mathrm{CO_2}$ atmosphere observed today. Recent work by \citet{Way2020} suggests that habitable conditions could have been maintained by the cloud albedo effect \citep{Yang2014} until planetary-scale volcanic events turned the climate into a hothouse. Alternatively, Venus could have been desiccated right after formation by the presence of a long-lived magma ocean that stayed melted for around 100 Myr \citep{Hamano2013}.

These pathways largely depend on the initial volatile inventory and their redistribution and loss throughout the planet's history. However, a planet that did undergo a runaway greenhouse effect with an Earth-like reservoir of water would in principle be able to keep its surface temperature up indefinitely until the water is completely lost to space. Runaway greenhouse theory indicates that the outgoing radiation fails to increase with increasing surface temperature until some threshold value, beyond which it increases drastically. What determines this threshold? One well-known mechanism is for the surface to become hot enough to start radiating to space directly through the window regions of the steam atmosphere. However, another mechanism that can increase the outgoing heat flux, less widely appreciated, is the penetration of the dry adiabat into the optically thin upper layers, which allows the planet to also radiate outside the window regions.

From a more general perspective, the observational bias towards extrasolar planets on close-in orbits implies that most detected rock-dominated exoplanets, such as super-Earths, are likely to reside within the instellation threshold of the runaway greenhouse \citep{2019AREPS..47..141J}. Such planets differ from temperate rocky planets in a variety of ways. 
Given a large enough water reservoir, the runaway greenhouse effect can inflate the atmosphere such that the increased radius would be indistinguishable from that of a planet with a larger radius free of a runaway state \citep{Turbet2019}. This inflation is caused by the additional water vapor relative to a temperate climate, which renders the atmosphere more opaque, raises the level of the infrared photosphere, cloud top, and overall scale height. The transit radii of runaway greenhouse planets can be further degenerate because of the different densities of molten versus solid rock \citep{Bower2019}.
From an evolutionary perspective, long-term runaway greenhouse states can desiccate the planet via H$_2$O photolysis in the upper atmosphere and subsequent hydrogen escape \citep{Kasting1987,Wordsworth2013,2015AsBio..15..119L,2016ApJ...829...63S} and substantially alter the planetary composition inherited from formation. The planetary radius can thus serve as a crucial indicator of possible surface environments, allowing for the determination of the steady-state surface temperature, the shape of the transmission spectrum of the atmosphere, or how much volatile content can be left in the planet's inventory above and below the surface.

In order to test theoretical predictions for rocky planets on both sides of the runaway greenhouse limit, astronomical surveys may allow for more accurate diagnoses of possible climate states and compositions of rock-dominated exoplanets and their atmospheres. To do so, measurements from indirect surveys with instruments such as PLATO \citep{2014ExA....38..249R}, ARIEL \citep{Tinetti2018} and JWST \citep{2019BAAS...51c..58B} to measure radii, and ESPRESSO \citep{Pepe2014} or SPIRou \citep{Artigau2014} to obtain mass measurements, or direct imaging techniques \citep{2019arXiv190801316Q,Quanz2021}, need to be combined and compared to theoretical models. In particular, the emission spectrum of the atmosphere and its radiating layer have a significant impact on long-term climate evolution and on potentially diagnostic characteristics of the atmospheric structure. Previous studies \citep{2016ApJ...829...63S,2017JGRE..122.1539M,Katyal20,Lichtenberg21JGRP} investigated how a runaway greenhouse effect can affect the spectral signature of the atmosphere.

This paper has three goals:
\begin{itemize}
    \item To develop and validate the \textsc{socrates} real-gas radiative transfer model for use in runaway and post-runaway climate calculations
    \item To develop a computationally efficient band-grey model suitable for rapid calculations with three dimensional global circulation models (GCMs), and evaluate the extent to which it reproduces the key features of the \textsc{socrates} calculation and other comprehensive radiative calculations that have appeared in the literature
    \item To better elucidate the factors that allow a planet which has undergone a runaway greenhouse to eventually increase its radiative cooling to space, thus reaching a new equilibrium surface temperature, and to determine the dependence of the post-runaway temperature on the initial water inventory. 
\end{itemize}
In pursuit of the third of these goals, we make use of a contribution function to determine the contribution of various levels in the atmosphere to the radiative emission to space. We show that the transition from a runaway to a post-runaway climate is marked in the infrared regions by a shift in the altitude of the radiating layer, here called the infrared photosphere of the planet, and in the visible and ultraviolet regions by a significant contribution of thermally emitted outgoing shortwave radiation. 

In Sect. \ref{sec:model} we describe our numerical methods to model runaway greenhouse climates and present their results in Sect. \ref{sec:results}. We discuss the significance of our results for exoplanet climates and astronomical surveys in Sect.  \ref{sec:discussion} and conclude in Sect.  \ref{sec:conclusion}.

\section{Methods} \label{sec:model}

Determination of the climate evolution requires models of the vertical transport of energy by radiation and convection.  In this paper we consider only clear-sky conditions in the radiative calculation. Though cloud radiative effects are certainly important in determining the conditions for a runaway, and the nature of the post-runaway climate, restriction to clear-sky conditions allows us to focus on the underlying physics of the runaway, and makes it easier to compare with existing results in the literature.  Furthermore, it is impossible to properly project cloud properties in 1D models, since the necessary dynamical information is not available in 1D; such processes are best treated in the context of 3D modeling.

For radiative transfer incorporating comprehensive real-gas opacity variations, we employ the \textsc{socrates} radiative transfer suite \citep{edwards1996socrates}, which makes use of the correlated-$k$ method to approximate the transmission function and solves the plane-parallel, two-stream approximated radiative transfer equations. For the \textsc{socrates} simulations, the line-by-line opacity data was taken from \cite{HITRAN2016}, choosing the main isotopologue $\mathrm{H_2^{16}O}$, and taking the complete available wavenumber range [0.072,25710.825] $\mathrm{cm}^{-1}$. The self-broadened continuum of water vapor was taken from \url{https://github.com/DavidSAmundsen/socrates_tools/tree/master/continuum/h2o}, using the files mt\_ckd\_v3.0\_s260 and mt\_ckd\_v3.0\_s296. 
Though \textsc{socrates} has the capability to represent scattering, our simulations ignore this effect for thermally emitted radiation, though we shall see eventually that in some cases there is enough thermal emission of shortwave radiation to make scattering potentially significant. 

The \textsc{socrates} suite uses the two-stream form of the Schwarzschild equations. In the absence of scattering, the up- and downward fluxes satisfy

\begin{eqnarray}
\frac{1}{D}\frac{d}{d\tau_\nu}I_+(\tau_\nu,\nu)&=&-I_+(\tau_\nu,\nu)+\pi B(\nu,T(\tau_\nu)), \label{eqn:Schwarzschild_Iplus} \\
\frac{1}{D}\frac{d}{d\tau_\nu}I_-(\tau_\nu,\nu)&=&I_-(\tau_\nu,\nu)-\pi B(\nu,T(\tau_\nu)), \label{eqn:Schwarzschild_Iminus}
\end{eqnarray}

with $D = 1.66$ the diffusivity factor approximating the angular dependence, $I_+$ and $I_-$ the upward and downward fluxes evaluated at the optical depth $\tau_\nu$ and at the frequency $\nu$, and $B$ the Planck function evaluated at the frequency $\nu$ and at the temperature $T(\tau_\nu)$.

To provide a simpler route to understanding the physical processes involved in runaway and post-runaway atmospheres, and to allow for the greater computational efficiency needed to facilitate later 3D calculations, we also introduce a simplified band-grey radiative transfer model.

Directly integrating the Schwarzschild integrands in an optically thick setting requires a very high vertical resolution, which makes getting a good accuracy difficult in a reasonable amount of time. In order to bypass this issue, we use an approximation using a two-stream source function method from \cite{https://doi.org/10.1029/JD094iD13p16287}, which is faster by more than one order of magnitude, well-suited for optically thick atmospheres, and most accurate in the limit of pure absorption. The source function technique is a two-stream method that uses ‘sources’ in the atmosphere (i.e. the blackbody flux), and follows the stream of the sources in the upward and downward directions, as a solution to the Schwarzschild equations. Instead of approximating the angular dependence with the diffusion coefficient mentioned above, we use a 2-point Gaussian quadrature with weights determined by the Gauss-Legendre quadrature law. In this formulation, the flux $I$ is integrated over all cosine angles.

\begin{equation}\label{eqn:Prequadrature}
I(\tau_\nu,\nu)=\pi \int\limits_{\mu}I_\mu d\mu,
\end{equation}

which in Gauss quadrature becomes

\begin{equation}\label{eqn:Quadrature}
I(\tau_\nu,\nu)=\pi \sum\limits_{\mu}w_\mu I_\mu \mu, 
\end{equation}

where $I_\mu$ is the directional intensity, $w_\mu$ is the weight for that quadrature point, and $\mu$ the cosine angle for that quadrature point.

Incorporation of two-stream scattering effects coupling the upward and downward streams is straightforward but will not be pursued here. 

The band-grey approximation involves dividing the spectral range into a small number of regions with opacity coefficients assumed to be uniform over each region. This allows a bandwise decomposition of any quantity computed by the model, a feature that we will use in Section \ref{sec:results}. Here we approximate two of the main near-infrared (NIR) atmospheric windows of water vapor, designated as W1 and W2 in this work. Their opacities are computed by taking the average between the Rosseland mean and the Planck mean of the absorption coefficients over each window. The Rosseland mean, which overweighs the smaller opacities, is defined as

\begin{equation}\label{eqn:Rosseland}
\frac{1}{\kappa_\mathrm{R}} = \left(\int\limits_{0}^{\infty} \frac{\partial B_\mathrm{\nu}}{\partial T}d\nu\right)^{-1}\int\limits_{0}^{\infty}\frac{1}{\kappa_\mathrm{\nu}} \frac{\partial B_\mathrm{\nu}}{\partial T}d\nu,
\end{equation}

whereas the Planck mean, which overweighs the larger opacities, is 

\begin{equation}\label{eqn:Planckmean}
\kappa_\mathrm{P} = \left(\int\limits_{0}^{\infty} B_\mathrm{\nu}d\nu\right)^{-1}\int\limits_{0}^{\infty}\kappa_\mathrm{\nu} B_\mathrm{\nu}d\nu,
\end{equation}

where $\kappa_\mathrm{R}$ is the Rosseland mean opacity, $\kappa_\mathrm{P}$ is the Planck mean opacity, $B_\nu$ is the Planck function depending on frequency $\nu$ and temperature $T$, and $\kappa_\nu$ is the opacity coefficient in $\mathrm{m}^2 \, \mathrm{kg}^{-2}$ corresponding to the frequency $\nu$.

\begin{table}[tbh]
\centering
\begin{tabular}{llr}
%\hline
Region    & Spectral range [$\mathrm{cm}^{-1}$]        & Opacity source \\ \hline
UV   & $(25000,40500)$     & \textsc{pokazatel} \\
VIS1 & $(18750,25000)$     & \textsc{pokazatel} \\
VIS2 & $(12500,18750)$     & \textsc{pokazatel} \\
IR   & $(1,12500)$ $-$ W1 $-$ W2     & \textsc{pokazatel}  \\
W1   & $(2200,2900)$    & \textsc{mt$\_$ckd} 3.4 \\
W2   & $(500,1300)$    & \textsc{mt$\_$ckd} 3.4 \\ \hline
$\textsc{socrates}$ & $(1, 24500)$ & \textsc{hitran}/\textsc{mt$\_$ckd} 3.4
\end{tabular}
\caption{Spectral regions and opacity sources considered in the band-grey model and $\textsc{socrates}$. The longwave bound of the band-grey spectral range, 1 $\mathrm{cm}^{-1}$, is the $\textsc{hitran}$ longwave bound. Its shortwave bound, 40500 $\mathrm{cm}^{-1}$, is the $\textsc{pokazatel}$ shortwave bound. When using a single shortwave band in the band-grey model, we combine UV, VIS1, and VIS2 into one band called SW going from 12500 to 40500 $\mathrm{cm}^{-1}$. When using two, we combine VIS1 and VIS2 into a single visible band, VIS, defined between 12500 and 25000 $\mathrm{cm}^{-1}$.}
\label{tab:1}
\end{table}

The \textsc{mt\_ckd} 3.4 data\footnote{\url{https://github.com/AER-RC/mt-ckd}} is given at a reference pressure and temperature of 1013 mbar and 260 K. It scales as the normalized density $\frac{\rho}{\rho_0} = \frac{P}{P_0}\frac{T_0}{T}$. The infrared and shortwave spectral regions outside the windows have their opacity averaged with the Rosseland and Planck means as above. The opacity data of these regions is taken from the ExoMol \textsc{pokazatel}\footnote{\url{https://chaldene.unibe.ch/data/Opacity3/1H2-16O__POKAZATEL_e0DAT} \citep{2015ApJ...808..182G}} water line list \citep{Exomol2020}, which is a theoretically derived line list reliable for temperatures up to and above 3000 K. An updated version of it will form the basis of the new HITEMP water line list at short wavelengths. For each of the four spectral regions defined in the band-grey model, we compute a mean opacity table that depends on pressure as well as temperature. The complete spectral range can be seen on Fig. \ref{fig:spectral_range_moreSWbs} and Panel A of Fig. \ref{fig:spectral_range_1SWb}, only at a pressure of 0.1 bar for clarity. The spectral range used in \textsc{socrates} as well as in each region included in the band-grey model are listed in Tab. \ref{tab:1} along with the sources of the corresponding opacity coefficients. We employ a total of 318 bands between the bounds of the \textsc{socrates} spectral range indicated in Tab. \ref{tab:1}.

At the high temperatures considered here, there is significant diffuse thermal emission in the visible (VIS) and ultraviolet (UV) spectral regions. These are treated identically to the longer wavelength radiation.  In the general case it is also necessary to consider the absorption and scattering of the incident direct-beam radiation received from the star, but as will be seen shortly the approach to atmospheric structure adopted in this paper does not require the vertical profile of the resulting fluxes, but only a specification of the net stellar energy absorbed by the planet, which is then balanced against the outgoing thermal planetary radiation. 

Any radiative transfer calculation requires the specification of the temperature-pressure profile of the atmosphere, as well as the surface temperature; for a clear-sky pure steam atmosphere, no other atmospheric conditions need to be specified.  As is common in runaway greenhouse calculations, we assume that the atmosphere has a deep convective layer extending upwards from the ground, within which the $T(p)$ profile is given by the appropriate adiabat. This assumption requires that there be sufficient heating of the atmosphere from below to trigger convection, but a water vapour atmosphere of a few bars or more is so optically thick that it takes very little flux (either from instellation reaching the ground or from geothermal flux) to maintain convection. In the post runaway state, there is no moisture source at the ground, so the atmosphere follows the dry water vapour adiabat until temperatures fall sufficiently that condensation occurs, whereafter the temperature follows the "dew point adiabat" obtained by solving Clausius-Clapeyron for temperature as a function of pressure.
 
In the unsaturated region extending upward from the planet's dry surface, the atmosphere is on the dry adiabat for water vapour, with corresponding lapse rate 

\begin{equation}\label{eqn:dry_adiabat}
\frac{d\ln T}{d\ln p} = \frac{R}{c_p(T)},
\end{equation}

which can be integrated numerically to obtain the profile $T(p)$. $R$ is the gas constant for water vapour and the temperature dependent specific heat $c_p(T)$ is taken from the NIST Chemistry WebBook, SRD 69 \citep{cox1984codata,chase1998nist}. It is assumed that turbulent fluxes keep the temperature of the air near the surface equal to the surface temperature $T_s$, which provides the boundary condition needed to integrate the equation. 

Within the saturated region aloft, the dew-point adiabat is given by

\begin{equation}\label{eqn:moist_adiabat}
T_{\mathrm{dew}}(p)=\frac{T_\mathrm{b}}{1-\frac{RT_\mathrm{b}}{L}\text{ln}\frac{p}{p_\mathrm{b}}},
\end{equation}

with $(T_\mathrm{b},p_\mathrm{b})$ the boiling point of water, or any point located on the same vapour/liquid co-existence curve as the boiling point. We have neglected the temperature dependence of latent heat in this formula and in all our calculations. Because we account for the temperature dependence of $c_p$, the code uses the same numerical procedure as \cite{Lichtenberg21JGRP}. A Runge-Kutta 4 integrator computes the dry pseudo-adiabatic $T(p)$ profile upward from the surface, up to the point where $T(p)<T_{\mathrm{dew}}(p)$, whereafter the slope changes from $\frac{R}{c_p}$ to $\frac{RT}{L}$ as per the Clausius-Clapeyron relation.  

On an adiabat, the temperature falls monotonically as the pressure is reduced to zero towards the top of the atmosphere, though on the dew-point adiabat this occurs very slowly owing to the logarithmic term in the denominator of Eq. \ref{eqn:moist_adiabat} and the small value of the prefactor multiplying the logarithm. In reality, the temperature structure of the upper atmosphere will be altered by the formation of a stratosphere, which is in pure radiative equilibrium.  The formation of a stratosphere warms the upper atmosphere and moderately increases radiation to space. In simplified 1D climate modeling it is common to attempt to represent this effect through inclusion of an isothermal stratosphere with temperature equal to the grey skin temperature \citep[e.g.][]{2013ApJ...765..131K}. \cite{Kasting1987,KASTING19911} adopted the same approach to climate equilibration with an overlying isothermal stratosphere in the case of early Venus and early Mars. The dew-point adiabat for water already keeps the upper atmosphere quite warm, and there is no indication that the inclusion of a stratosphere has an important effect on the results for thick steam atmospheres. The neglect of the stratosphere leads to an underestimation of the outgoing planetary radiation to some extent, though it is likely to be slight for the steam atmospheres considered here. In any event, the isothermal assumption yields a poor representation of actual stratospheric behaviour for non-grey atmospheres, and the use of the grey skin temperature is problematic for atmospheres with significant wavelength-dependent opacity.  If the stratosphere is thought to be important, it needs to be treated with a full radiative-convective formulation allowing for the formation of one or more pure-radiative layers, using forward calculations, either time-stepped or otherwise iterated toward solutions. Here, to allow for a focus on the key phenomena determining post-runaway climate, we adopt the simpler all-troposphere model of atmospheric structure, which does not require iterating. 

Because the atmospheric $T(p)$ structure in this approximation can be written solely in terms of the surface temperature and pressure, the distribution of heating from atmospheric absorption of the incoming stellar flux does not enter into the calculation. The stellar heating affects the amount of convective heat transport needed to maintain the adiabat, but not the atmospheric structure or outgoing planetary radiation. Surface temperature is determined simply by balancing net top-of-atmosphere absorbed stellar radiation against outgoing planetary radiation, requiring only the specification of instellation and planetary albedo. In this paper, we only compute the outgoing planetary radiation as a function of the surface temperature, and do not embark on a calculation of the planetary albedo; we only use an assumed albedo to provide some representative surface temperatures, and the reader is free to substitute other preferred values of albedo if warranted, e.g., by different surface characteristics. The planetary albedo is thus assumed to be constant at 0.2 for surface temperatures above 500 K, based on the planetary albedo calculations of \cite{2013ApJ...765..131K}, \cite{Goldblatt2013}, and \cite{2015ApJ...806..216H}.

The flux contribution function to the outgoing planetary radiation, $\mathcal{CF}_\mathrm{F}$ as well as the surface flux contribution $\mathcal{SCF}_\mathrm{F}$ are derived in Appendix \ref{A1}. Their expressions are

\begin{equation}\label{eqn:cff_eq}
\mathcal{CF}_F = 2\pi \pi B(T(\tau_{12}))\frac{e^{-D(\tau_\infty - \tau_2)}-e^{-D(\tau_\infty - \tau_1)}}{D},
\end{equation}

and

\begin{equation}\label{eqn:apx11}
\mathcal{SCF}_F=I_+(0) e^{-\tau_\infty},
\end{equation}

with $B(T(\tau_{12}))$ the Planck function at the temperature of the layer between $\tau_{1}$ and $\tau_{2}$, $\tau_\infty$ the optical thickness of the atmospheric column, and $D$ the diffusivity factor defined in Eq. \ref{eqn:Schwarzschild_Iplus} and \ref{eqn:Schwarzschild_Iminus}.

\section{Results} \label{sec:results}

\begin{figure*}[bht]
\centering
\includegraphics[width=0.99\textwidth]{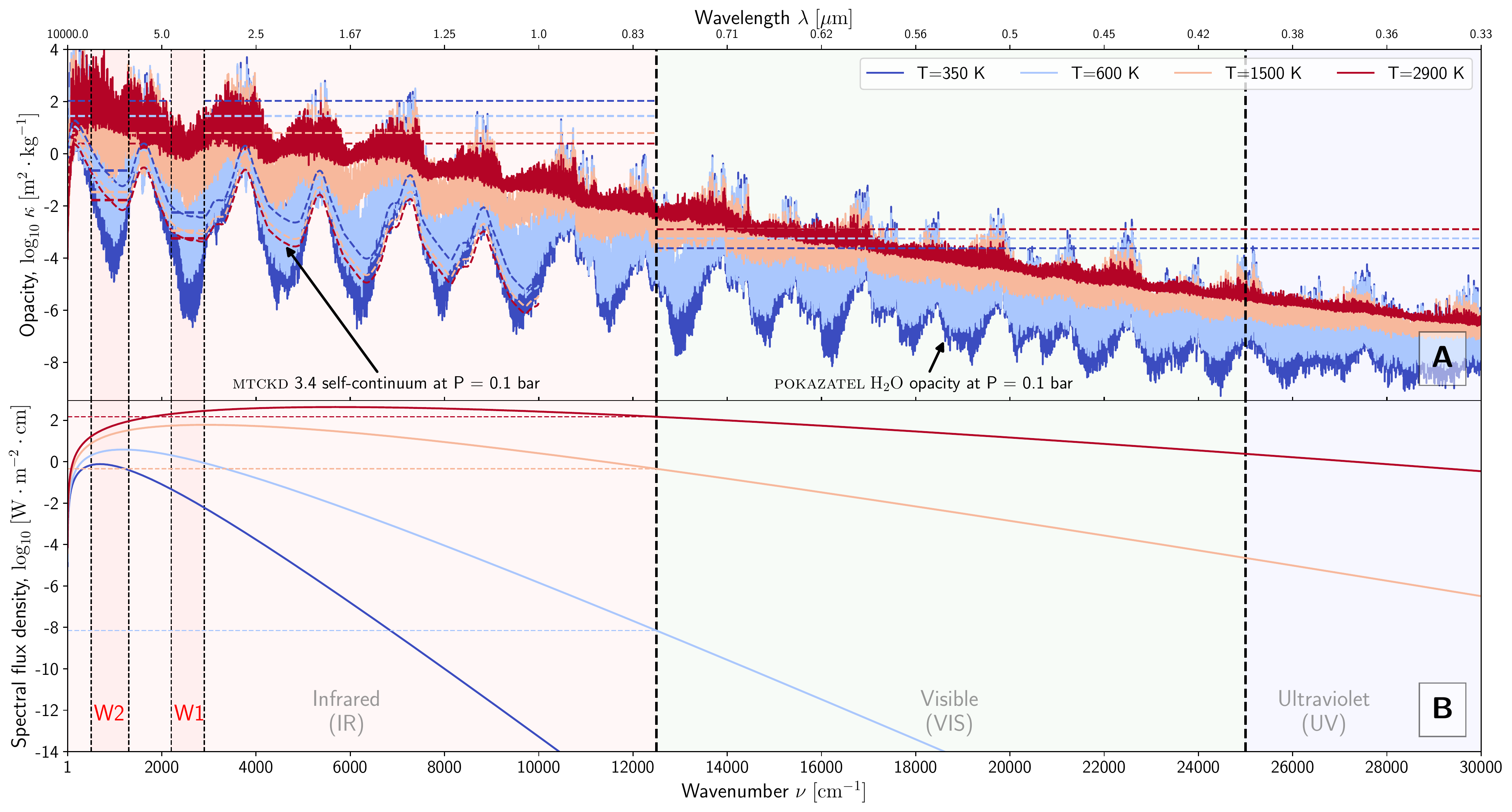} 
\caption{\textbf{(A)} Water vapor opacity data from the \textsc{pokazatel} line list for selected temperatures and a pressure of 0.1 bar. The \textsc{mt\_ckd} 3.4 self-broadened continuum has been superimposed between 1 and 10000 cm$^{-1}$ (dashed colored lines). Horizontal dashed lines of matching colors represent the mean opacity computed from taking the average between the Rosseland mean and the Planck mean in each specified spectral region, for each temperature. \textbf{(B)} Spectral radiance as a function of wavenumber for selected surface temperatures. Horizontal dashed lines mark the intersection of the Planck curves with the visible range of the spectrum. Temperatures higher than $\approx$2000 K yield significant visible and ultraviolet fluxes. Case with a single shortwave band.}
\label{fig:spectral_range_1SWb}
\end{figure*}
\begin{figure*}[bht]
\centering
\includegraphics[width=0.99\textwidth]{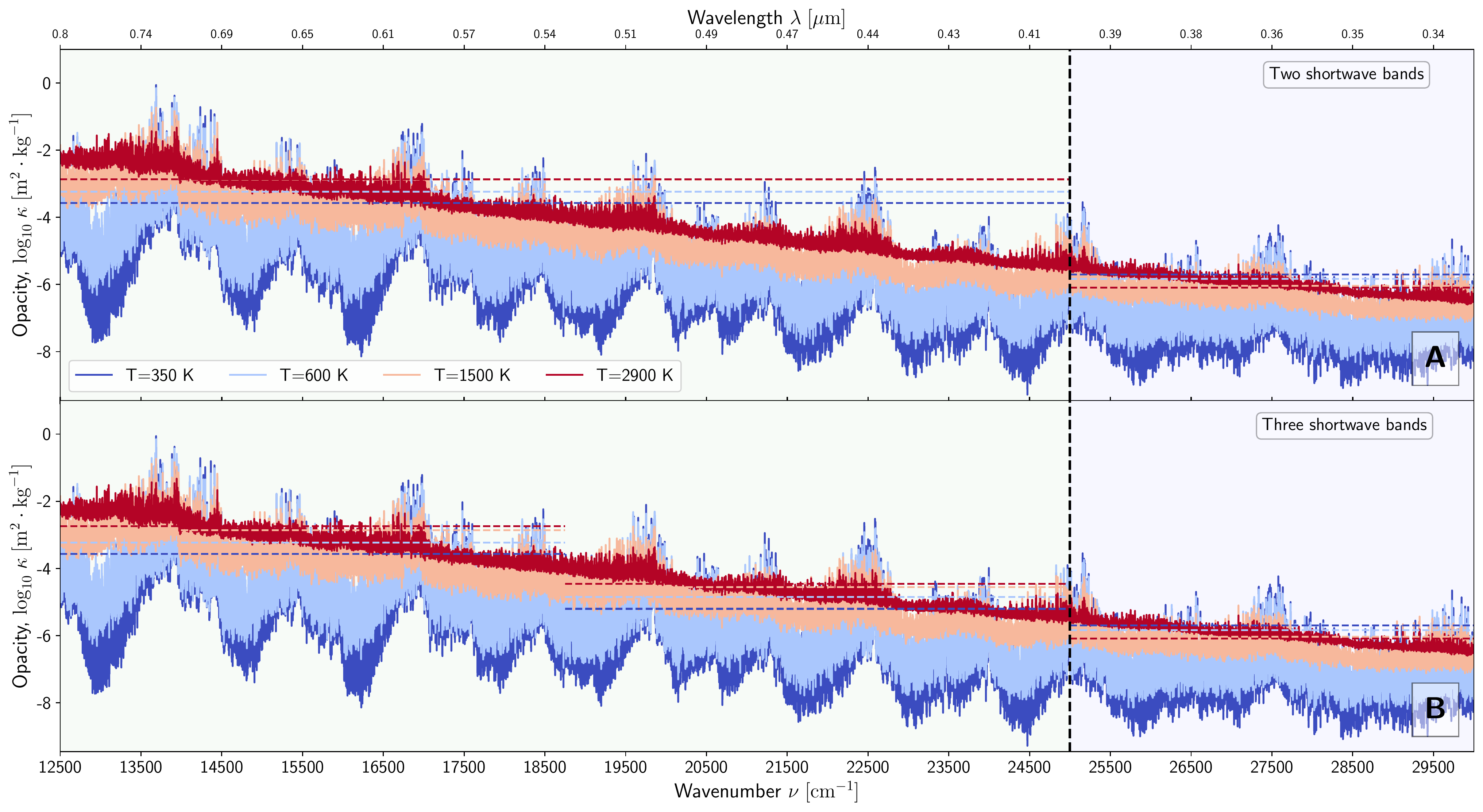} 
\caption{Same as Panel A of Fig. \ref{fig:spectral_range_1SWb}, but focusing on the shortwave part of the spectrum and displaying cases with \textbf{(A)}  two and \textbf{(B)} three shortwave bands.}
\label{fig:spectral_range_moreSWbs}
\end{figure*}

Fig. \ref{fig:spectral_range_moreSWbs} and Panel A of Fig. \ref{fig:spectral_range_1SWb} show the opacity coefficient of water vapor between 1 and 30000 $\mathrm{cm}^{-1}$, for 350, 600, 1500, and 2900 K, and for a pressure of 0.1 bar. The wavenumber range has been cut at 30000 $\mathrm{cm}^{-1}$ for clarity. The calculations of the band-grey model take into account the shortwave regions up to 40500 $\mathrm{cm}^{-1}$, while those of \textsc{socrates} stop at 24500 $\mathrm{cm}^{-1}$. The solid lines show the ExoMol \textsc{pokazatel} line list, and the dashed curves between 1 and 10000 $\mathrm{cm}^{-1}$ show the \textsc{mt\_ckd} 3.4 lines plus self-broadened continuum. Horizontal dashed lines represent the average of the Rosseland means and the Planck means computed over each region and for each temperature. They are used in the formation of the band-grey model. Figure \ref{fig:spectral_range_1SWb} shows the average as taken by the band-grey model when using a single shortwave band. Fig. \ref{fig:spectral_range_moreSWbs} shows how the average changes when the band-grey model uses two and three shortwave bands. Panel B of Figure \ref{fig:spectral_range_1SWb} shows the spectral flux density of a blackbody for each temperature. Intersections of the Planck curves with the beginning of the visible range of the spectrum are shown as horizontal dashed lines, which gives an impression of the contribution of shortwave regions with higher temperatures.

It is important to note that what is often called outgoing "longwave" radiation actually refers to the total thermally emitted radiation coming out of the planet, which includes the shortwave parts of the spectrum up to 40500 $\mathrm{cm}^{-1}$ in the case of the calculations presented in this work. Since we define explicit "infrared" and "shortwave" regions in our band-grey model, and because at the high temperatures we work with, the terms "longwave" and "shortwave" as they are often understood can be misleading, we will refer to this total outgoing radiation as Outgoing Planetary Radiation (OPR) in both models, meaning that the name will account for the provenance of the radiation instead of its place on the spectrum. Similarly, the incoming "shortwave" radiation will then be the Incoming Stellar Radiation (ISR).
\vspace{0.1cm}
\subsection{Photosphere shift} \label{sec:LW_cooling}

\begin{figure*}[tbh!]
\centering
\includegraphics[width=.99\textwidth]{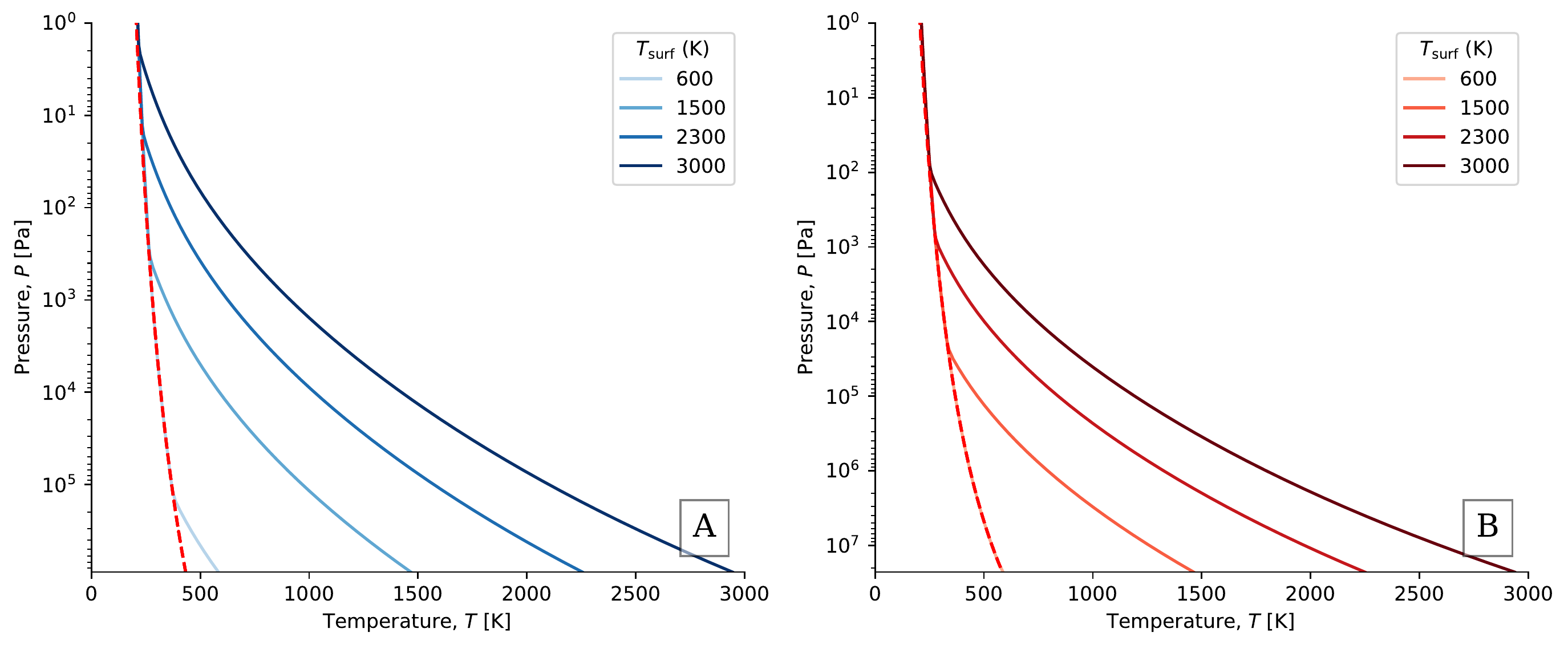}
\caption{Temperature profiles used in the band-grey model and in \textsc{socrates} for selected surface temperatures, at a surface pressure of \textbf{(A)} 10 and \textbf{(B)} 260 bars. The dashed red line is the Clausius-Clapeyron curve of water vapor.}
\label{fig:PT}
\end{figure*}

\begin{figure*}[tbh!]
\centering
\includegraphics[width=.99\textwidth]{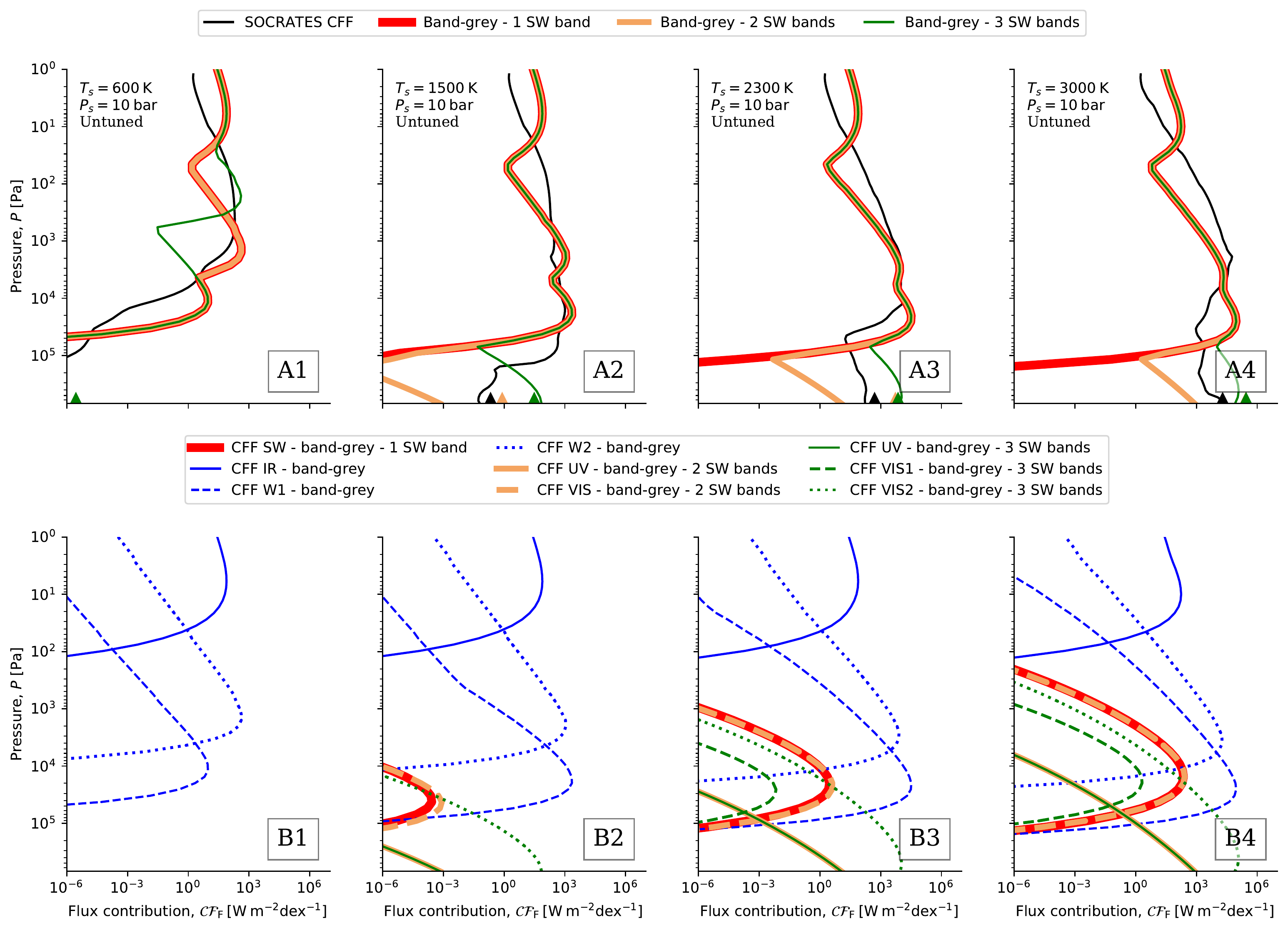}
\caption{Pressure as a function of the flux contribution function $\mathcal{CF}_\mathrm{F}$ to the OPR, for a surface temperature of \textbf{(A1/B1)} 600K, \textbf{(A2/B2)} 1500K, \textbf{(A3/B3)} 2300K, \textbf{(A4/B4)} 3000K. The $\mathcal{SCF_F}$ is shown as triangle markers. On the first row, the $\mathcal{CF}_\mathrm{F}$ is summed over all bands. On the second row, the band-grey $\mathcal{CF}_\mathrm{F}$ is decomposed bandwise and the \textsc{socrates} curve is not included. Cases using one, two, and three shortwave bands are compared. The surface pressure is 10 bar and the band-grey shortwave opacities are not tuned.}
\label{fig:EVSFM10_baseline}
\end{figure*}

\begin{figure*}[tbh!]
\centering
\includegraphics[width=.99\textwidth]{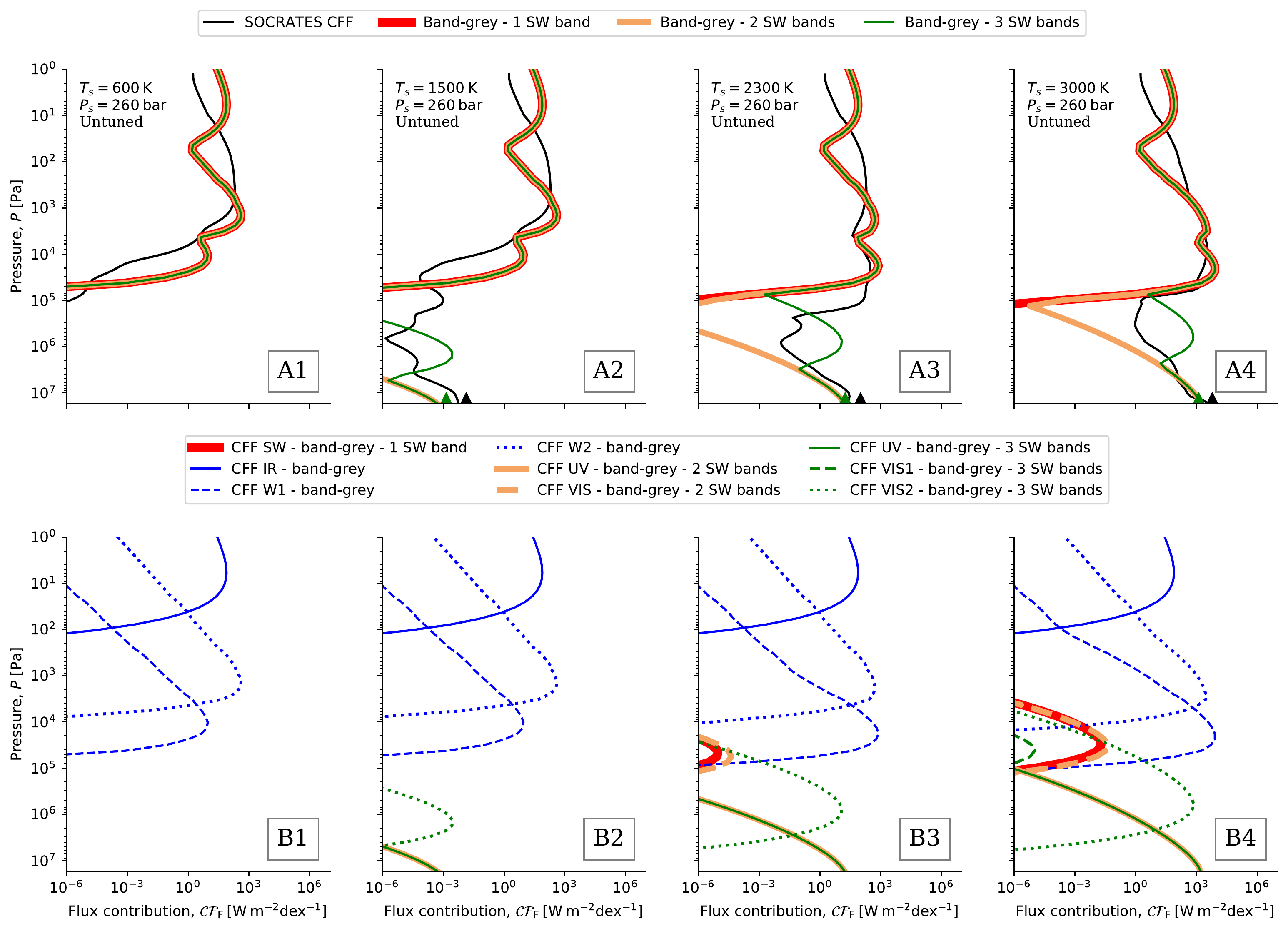}
\caption{Same as Fig. \ref{fig:EVSFM10_baseline}, but in this case the surface pressure is 260 bar and the band-grey shortwave opacities are not tuned.}
\label{fig:EVSFM260_baseline}
\end{figure*}

\begin{figure*}[tbh!]
\centering
\includegraphics[width=.99\textwidth]{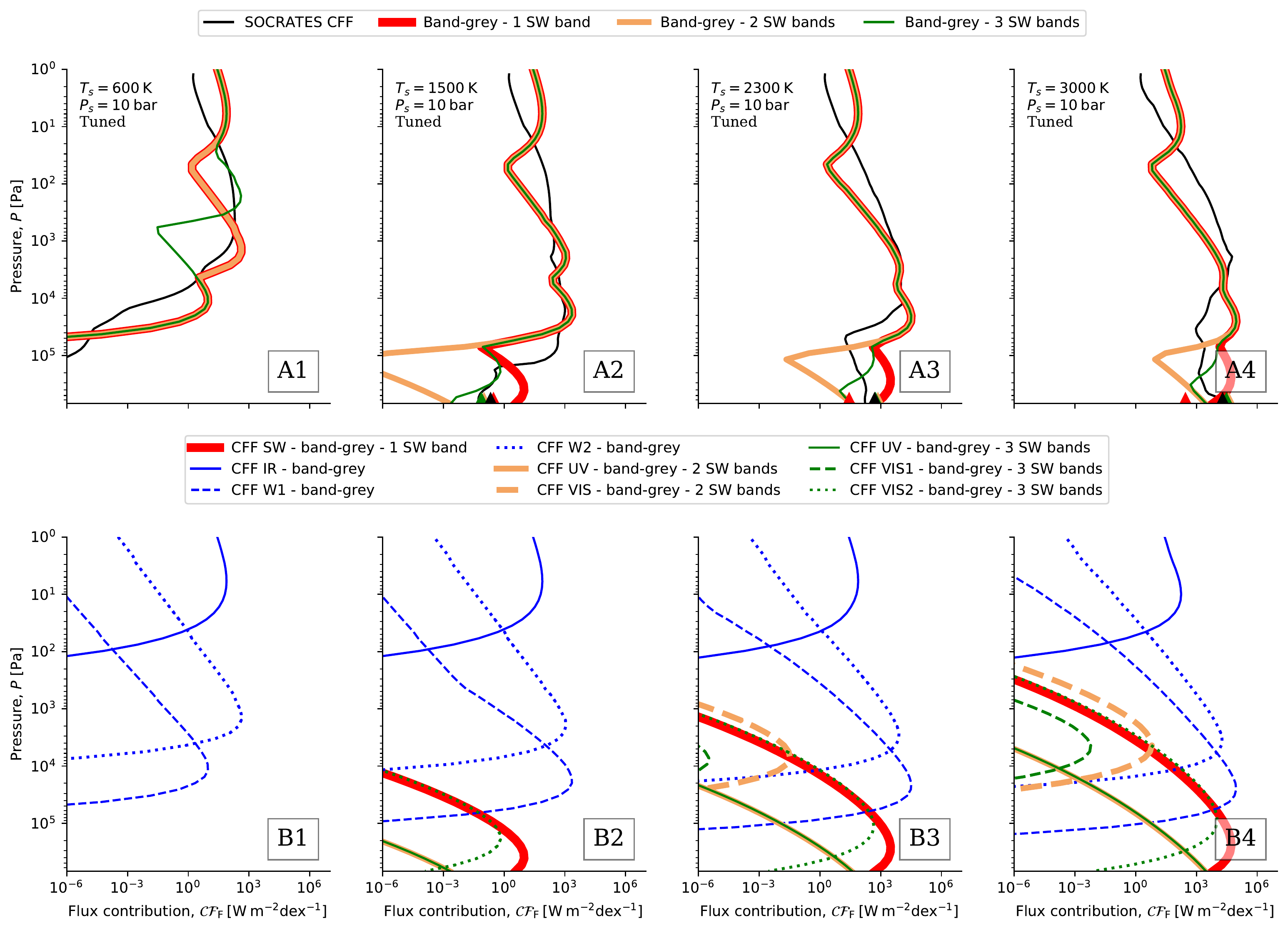}
\caption{Same as Fig. \ref{fig:EVSFM10_baseline}, but in this case the surface pressure is 10 bar and the band-grey shortwave opacities are tuned to the \textsc{socrates} outgoing planetary radiation.}
\label{fig:EVSFM10_tuned}
\end{figure*}

\begin{figure*}[tbh!]
\centering
\includegraphics[width=.99\textwidth]{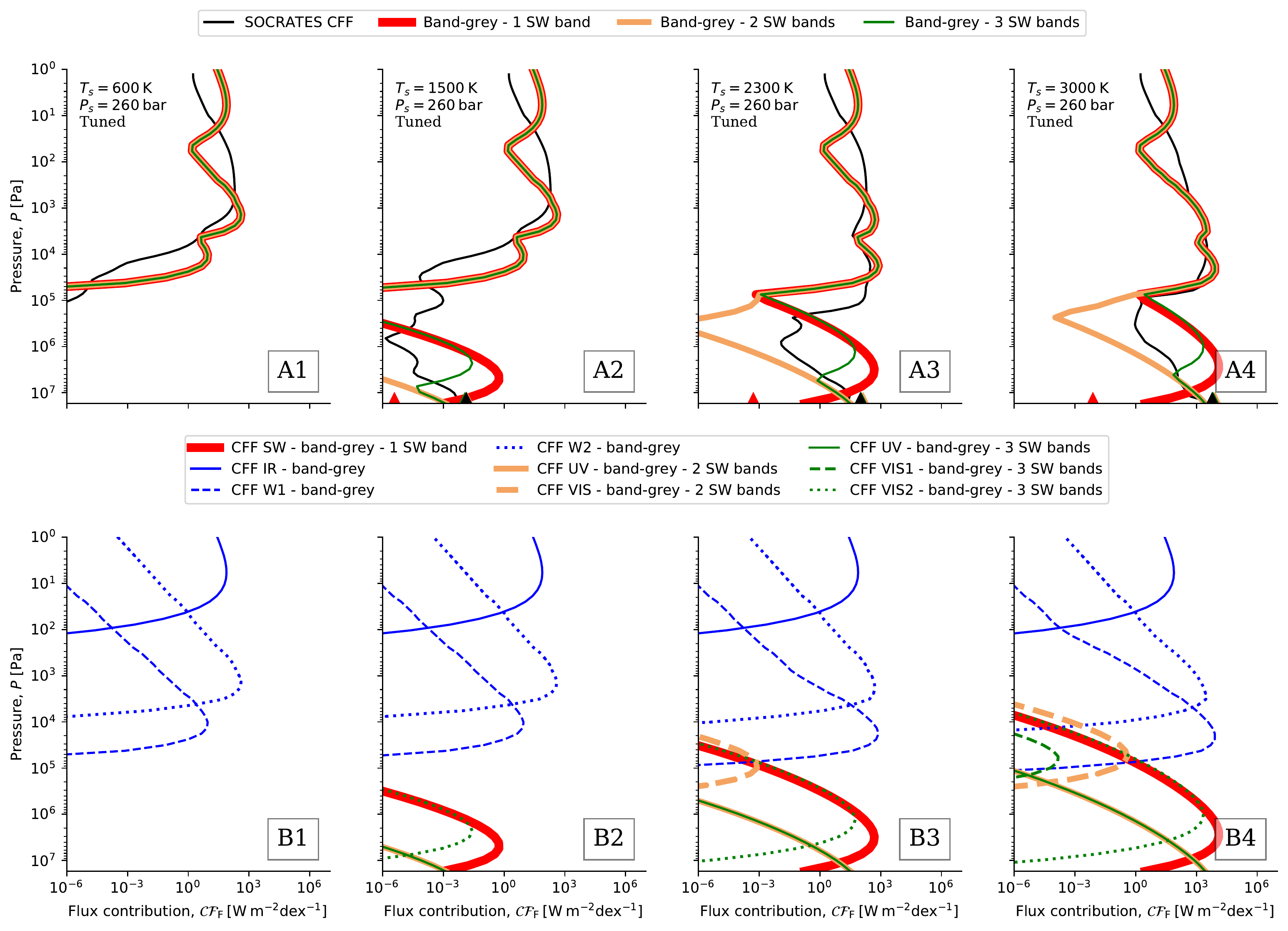}
\caption{Same as Fig. \ref{fig:EVSFM10_baseline}, but in this case the surface pressure is 260 bar and the band-grey shortwave opacities are tuned to the \textsc{socrates} outgoing planetary radiation.}
\label{fig:EVSFM260_tuned}
\end{figure*}

\begin{figure}[tb]
\centering
\includegraphics[width=0.49\textwidth]{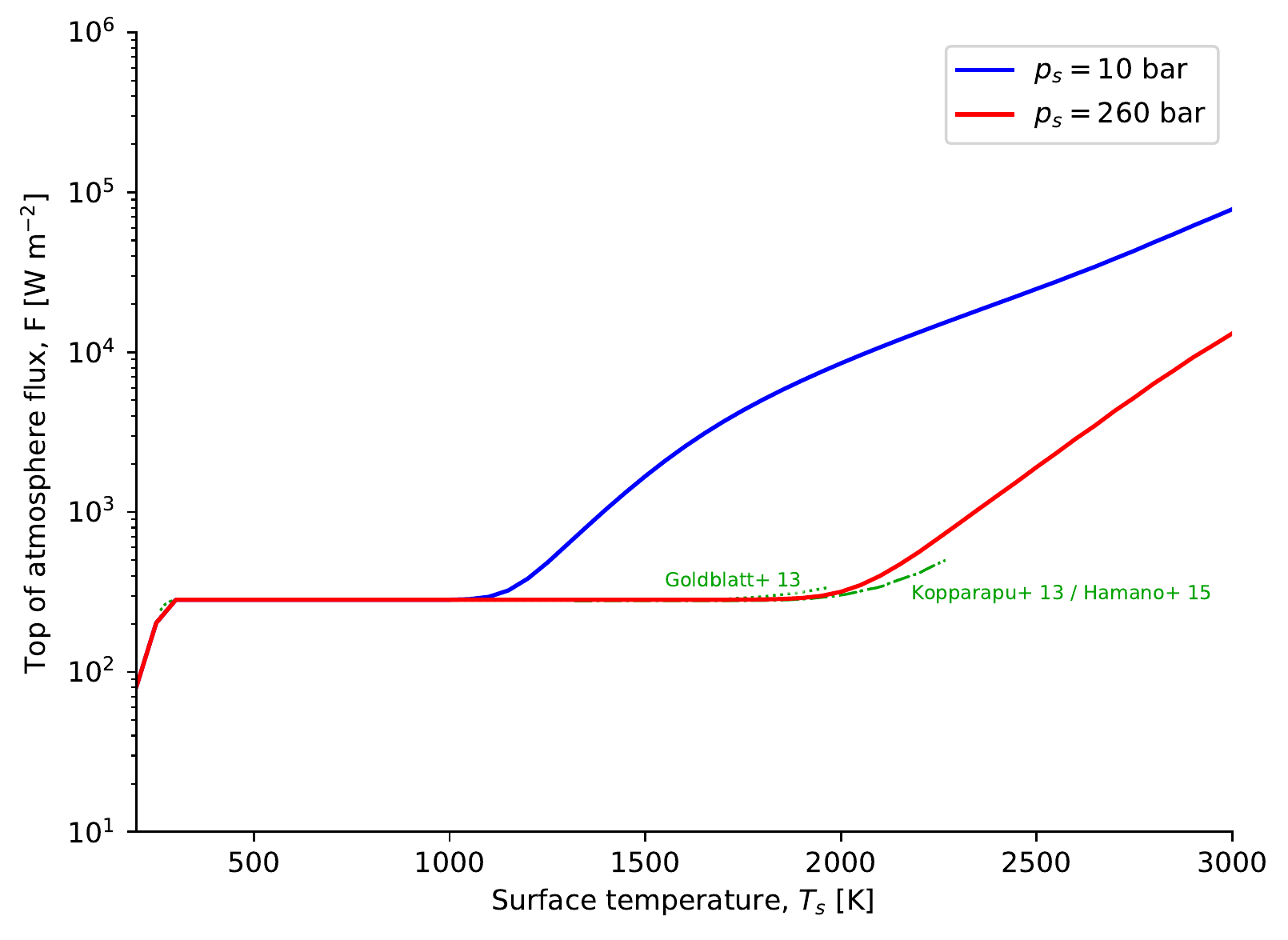}
\caption{Outgoing planetary radiation obtained with \textsc{socrates} for a surface pressure of 10 and 260 bars. Included reference curves from \citet{2013ApJ...765..131K}, \citet{Goldblatt2013}, and \citet{2015ApJ...806..216H}.}
\label{fig:soc_olr}
\end{figure}
\begin{figure}[tb]
\centering
\includegraphics[width=0.49\textwidth]{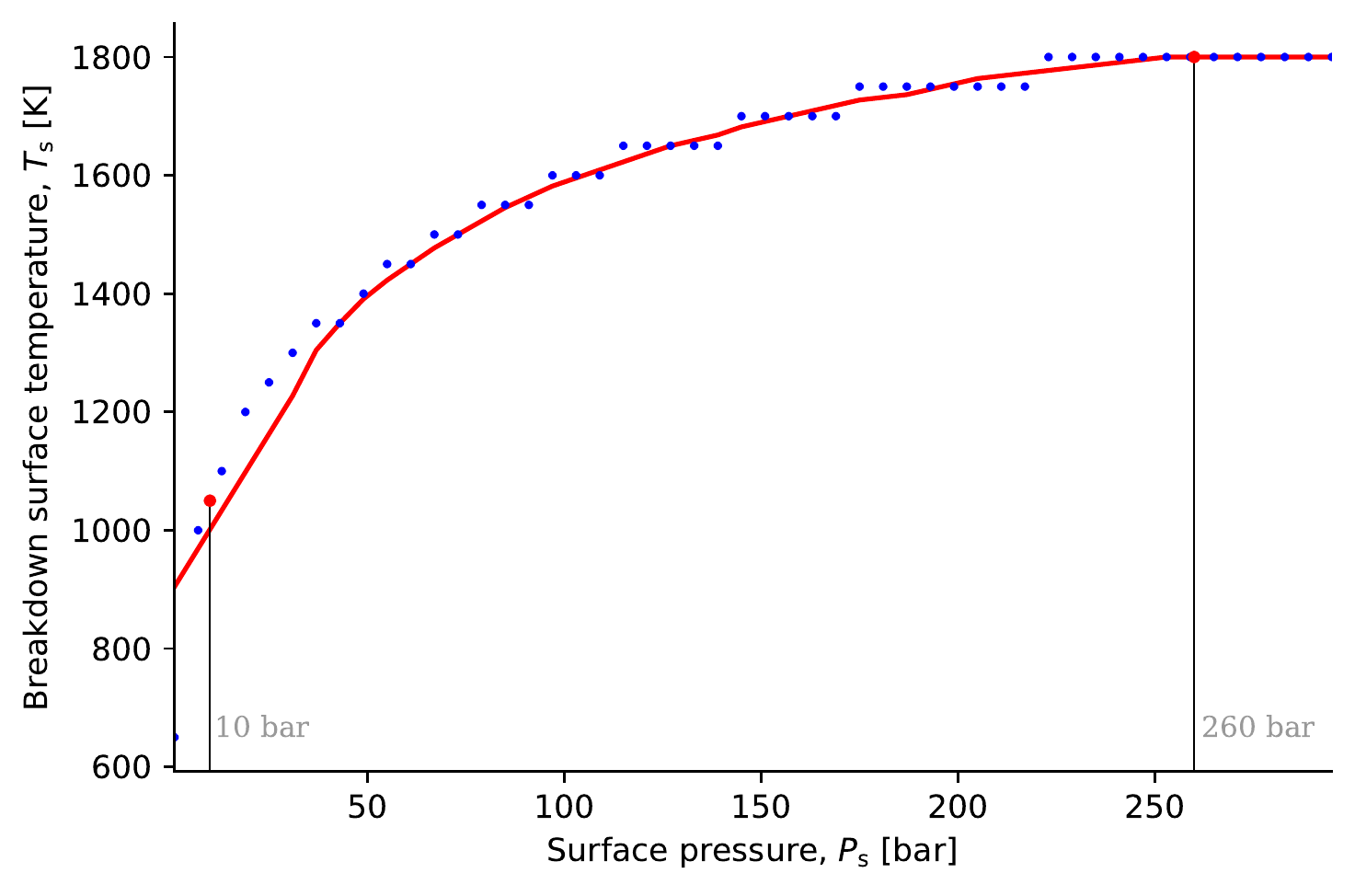}
\caption{Surface temperature at which the OPR increases by 1 $\mathrm{W} \, \mathrm{m}^{-2}$, marking the end of the tropospheric radiation limit, as a function of surface pressure. Surface pressures go from 1 to 295 bars, with a step of 6 bars. The discrete data has been smoothed, the result of it showed by a red line. Red dots show the positions of the reference 10 and 260 bar surface pressures.}
\label{fig:ps_Ts}
\end{figure}
Figures \ref{fig:EVSFM10_baseline} to \ref{fig:EVSFM260_tuned} shows the results of simulations performed with the \textsc{socrates} suite and with the band-grey model, in the pure steam limit. A surface pressure of 260 bar was chosen to reproduce the curves obtained by \citet{2013ApJ...765..131K}, \citet{Goldblatt2013}, and \citet{2015ApJ...806..216H}, visible on Figures \ref{fig:soc_olr}, \ref{fig:OLR_OSR_baseline}, and \ref{fig:OLR_OSR_tuned}. We added a surface pressure of 10 bar to provide a sense of the impact of surface pressure on outgoing radiation. A surface pressure of 10 bars in a post-runaway atmosphere corresponds to a planet with a limited water inventory (about 5\% of Earth's), or a planet that has lost most of a more massive ocean through photolysis, hydrogen escape, and geochemical sinks of oxygen. 

Figures \ref{fig:PT}A and B show the temperature profiles from which fluxes were computed, for four selected surface temperatures of 600, 1500, 2300 and 3000 K, and for a surface pressure of 10 and 260 bar, respectively. The saturation vapor pressure curve for water vapor, $p_{\mathrm{sat},\mathrm{H_2O}}$, is shown as a red dashed line, superimposed by the thermal profile with the smallest surface temperature.

Figures \ref{fig:EVSFM10_baseline}, \ref{fig:EVSFM260_baseline}, \ref{fig:EVSFM10_tuned}, and \ref{fig:EVSFM260_tuned} show the profile of the flux contribution function to the outgoing planetary radiation, $\mathcal{CF}_\mathrm{F}$, along with the surface flux contribution $\mathcal{SCF}_\mathrm{F}$, for the same selection of surface temperatures and surface pressures. It computes how many watts per square meter each atmospheric layer contributes to the top of atmosphere \citep{Drummond2018}. The upper panels show the total $\mathcal{CF}_\mathrm{F}$, summed over all spectral bands. The blue curves in these panels show the results of simulations performed with the \textsc{socrates} suite in the pure steam limit. The corresponding band-grey simulation results are shown in red, orange, and green lines, when using a single shortwave band, two, or three, respectively. The $\mathcal{SCF}_\mathrm{F}$ is also plotted as triangles of the corresponding color. The lower panels show a bandwise decomposition of the $\mathcal{CF}_\mathrm{F}$ obtained with the band-grey model, which allows an easy distinction of $\mathcal{CF}_\mathrm{F}$ by spectral region.
Figures \ref{fig:EVSFM10_baseline} and \ref{fig:EVSFM260_baseline} show baseline cases, at surface pressures of 10 bar and 260 bar respectively, where the averaged shortwave opacity coefficients used in the band-grey model have not been tuned.
Figures \ref{fig:EVSFM10_tuned} and \ref{fig:EVSFM260_tuned} show cases where the shortwave opacity coefficients of the band-grey model have been manually tuned to match the outgoing planetary radiation obtained with the \textsc{socrates} suite. The tuning coefficients used in each band and at each surface pressure can be seen in Table \ref{tab:2}.

\begin{table}[tbh]
\centering
\begin{tabular}{llr}
%\hline
Region    & $p_s = 10$ bar & $p_s = 260$ bar \\ \hline
SW   & $4.6 \; 10^{-2}$     & $4.23 \; 10^{-3}$ \\
UV, VIS & $5$     & $4.9 \; 10^{-1}$ \\
UV, VIS1, VIS2 & $6$     & $5.4 \; 10^{-1}$ \\
IR   & $0.2$     & $0.2$  \\
W1   & $0.2$    & $0.2$ \\
W2   & $0.2$    & $0.2$
\end{tabular}
\caption{Tuning coefficients used in the band-grey model. When the shortwave region SW is divided into two (UV and VIS) or three (UV, VIS1, and VIS2) sub-regions, the tuning coefficients between them are the same.}
\label{tab:2}
\end{table}

At a surface temperature of 600 K, all three band-grey cases explored here follow the behavior that \textsc{socrates} yields, and the atmosphere at pressures higher than 1 bar does not contribute to the outgoing planetary radiation in any of the cases explored. The $\mathcal{CF}_\mathrm{F}$ increases with height and peaks at about 400-1000 Pa, which sits comfortably within the moist convective zone. The two windows of water vapor accounted for are more transparent, and therefore bring down the peak relative to what the IR region alone would yield. They also create a secondary peak at about 0.1 bar, due to the bluest window region, W1. Different spectral regions contribute to the OPR with peaks located at different levels because each region has a different width and opacity, and the Planck contribution varies in each of them. The cases with one, two, and three shortwave bands are superimposed as the shortwave contribution is negligible at this temperature. 

Going up to $T_\mathrm{s} = 1500$ K, the atmosphere already contributes a significant amount of energy at subsaturated levels at $p_s = 10$ bar, however at $p_s = 260$ bar, most of the contribution still comes from the moist convective zone, even in the tuned case where the single shortwave band (SW) contribution curve peaks at about 20 bar but stays below 1 $\mathrm{W} \, \mathrm{m}^{-2} \, \mathrm{dex^{-1}}$. The dex unit is adimensional; it is an order of magnitude. A difference of $x$ dex is a change by a factor $10^x$. The primary peak due to reddest window region, W2, is relegated to secondary peak, whereas the previous secondary peak becomes the most important. Both peaks do not move noticeably from the $T_\mathrm{s} = 1500$ K case, however. The shortwave contribution starts to show near the surface up to a millibar at $p_s = 10$ bar and up to 3 bar at $p_s = 260$ bar. It reaches the surface when we use two or three shortwave bands, but the case with a single shortwave band requires tuning to get the shortwave $\mathcal{CF}_\mathrm{F}$ to do the same. Using three shortwave bands yields a larger shortwave $\mathcal{CF}_\mathrm{F}$ near the ground than using two.

At $T_\mathrm{s} = 2300$ K and $3000$ K, the shortwave contribution near the ground becomes dominant, overcoming the contribution of the infrared region. This makes the contribution to the OPR come from both saturated and subsatured parts of the atmosphere. The contribution of the surface itself also becomes important at these high surface temperatures. 

Figure \ref{fig:soc_olr} shows the top of atmosphere energy budget including the outgoing planetary radiation and the incoming stellar radiation, as a function of the surface temperature and for both surface pressures explored on Figures \ref{fig:EVSFM10_baseline} to \ref{fig:EVSFM260_tuned}, 10 and 260 bars. 

After an initial increase, the OPR stays flat between 300 and 1800 K in the 260 bar case, and between 300 and 1050 K in the 10 bar case. This flattening represents the familiar limiting "OLR", called the Simpson-Nakajima limit by \cite{Goldblatt2013}. In this paper, we will refer to it as the 'limiting OPR', the 'OPR limit', or the 'tropospheric radiation limit', denoted by $\mathrm{OPR}_\infty$. It then increases at hotter surface temperatures, almost at an exponential rate. The 260 bar curve fits relatively well the reference curves shown in dotted and dashed green lines, showing that the \textsc{socrates} radiation code, which is widely used in planetary climate studies, can reproduce the OPR limit found in other calculations. The tropospheric radiation limit is about $\mathrm{OPR}_\infty =$ 283 $\mathrm{W} \, \mathrm{m}^{-2}$.

Figure \ref{fig:ps_Ts} shows how the surface temperature at which the outgoing planetary radiation limit breaks down evolves as a function of surface pressure. The blue points are \textsc{socrates} simulations computing the surface temperature at which $\mathrm{OPR}(T_\mathrm{s}) - \mathrm{OPR}_{\mathrm{ref}} > 1$, with OPR the Outgoing Planetary Radiation and $\mathrm{OPR}_{\mathrm{ref}}$ a reference OPR value arbitrarily chosen as $282.7 \: \mathrm{W} \, \mathrm{m}^{-2}$. The temperature at which the limit breaks down increases non-linearly with surface pressure and reaches a constant value of about $T_\mathrm{s,breakdown} = 1800$ K. Even though the OPR limit itself does not depend on the surface pressure, which in turn depends on the mass of the original ocean, the breakdown that allows the OPR to increase again strongly depends on the surface pressure, at least up to about 295 bars. The non-linearity of this increase comes from the non-linearity of the speed at which the intersecting point between the dry and moist adiabats rises, as well as from the non-linearity of the increase in contribution of the Planck function in the shorter waves, as the surface warms up.

\subsection{Shortwave contribution to outgoing radiation} \label{sec:SW_cooling}

\begin{figure*}[tbh!]
\centering
\includegraphics[width=.99\textwidth]{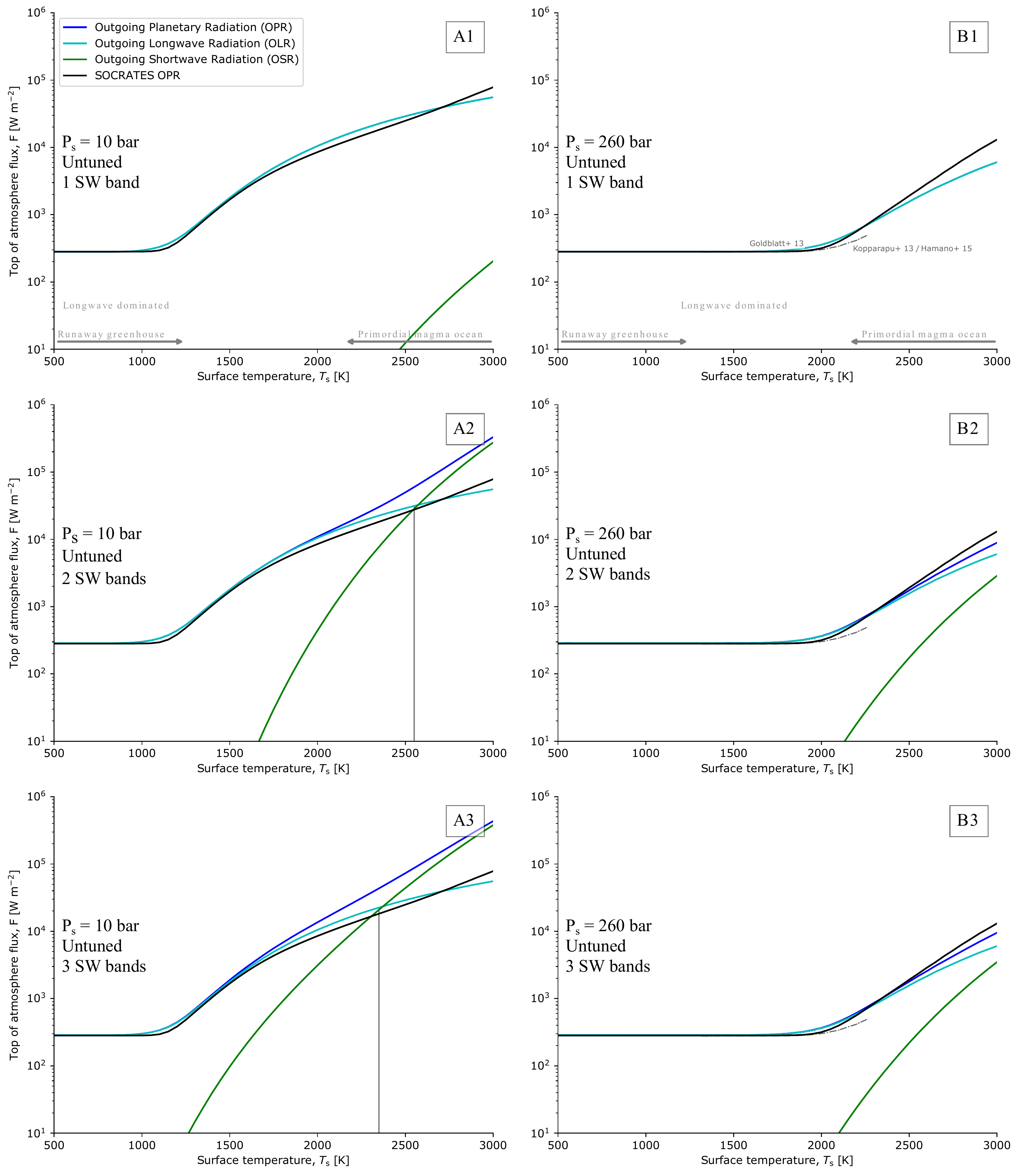}
\caption{Band-grey energy budget at the top of the atmosphere for a surface pressure of \textbf{(A)} 10 bars and \textbf{(B)} 260 bars in a pure steam atmosphere. The first row uses a single shortwave band, the second uses two, and the third uses three. The dark blue line is the  Outgoing Planetary Radiation (OPR), the cyan line is the Outgoing Longwave Radiation (OLR), the green line is the Outgoing Shortwave Radiation (OSR), and the black line is the OPR computed by \textsc{socrates}. The reference curves from \cite{2013ApJ...765..131K}, \cite{Goldblatt2013}, and \cite{2015ApJ...806..216H} are plotted in the 260 bar case in the right panels. A black vertical line divides the surface temperature domain into longwave- and shortwave-dominated parts wherever the OSR exceeds the OLR. The tropospheric radiation limit is found at about $\mathrm{OPR}_\infty = $ 281 $\mathrm{W} \, \mathrm{m}^{-2}$. The band-grey shortwave opacities are not tuned.}
\label{fig:OLR_OSR_baseline}
\end{figure*}

\begin{figure*}[tbh!]
\centering
\includegraphics[width=.99\textwidth]{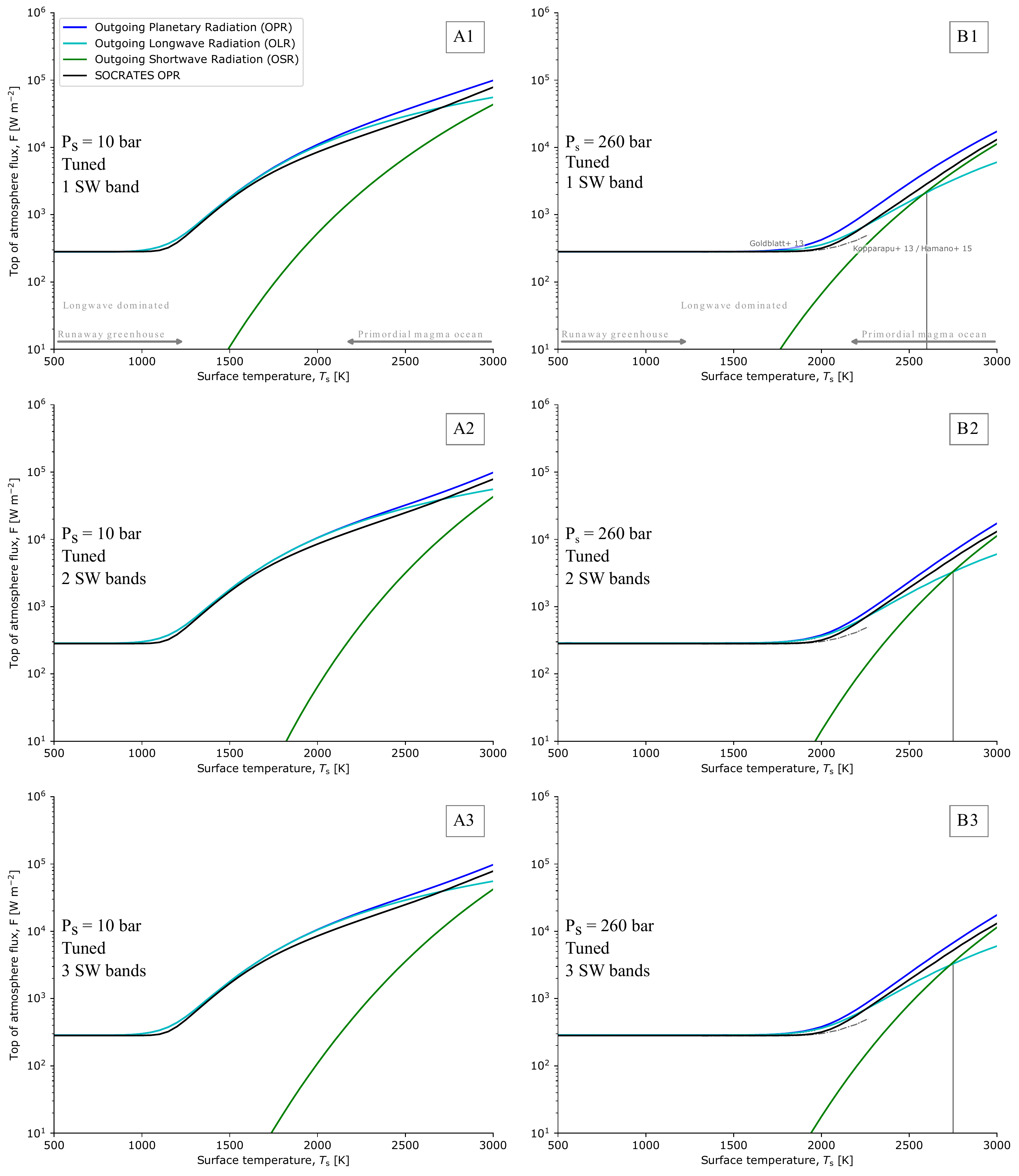}
\caption{Same as Fig. \ref{fig:OLR_OSR_baseline}, but here band-grey shortwave opacities are tuned to the \textsc{socrates} outgoing planetary radiation.}
\label{fig:OLR_OSR_tuned}
\end{figure*}

Figures \ref{fig:OLR_OSR_baseline} and \ref{fig:OLR_OSR_tuned} show results from pure-steam simulations using the band-grey model with the same parameters as in Figure \ref{fig:soc_olr}, for $p_\mathrm{s} =$ 10 and 260 bar respectively. They show the top of atmosphere radiation budget as a function of the surface temperature. It includes the outgoing planetary radiation in dark blue, which is the sum of the outgoing longwave radiation, plotting in cyan, that comes strictly from the IR, W1, and W2 regions of the spectrum, and the outgoing shortwave radiation, shown in green. In this work, outgoing shortwave radiation refers to thermal shortwave emission, not reflection. 
Figure \ref{fig:OLR_OSR_baseline} shows baseline band-grey simulations where the averaged shortwave opacities have not been tuned to the \textsc{socrates} outgoing planetary radiation. The first row of Figure \ref{fig:OLR_OSR_baseline} uses only one shortwave band, and the shortwave contribution is negligible at all surface temperatures. The second and third rows use two and three shortwave bands respectively. As we can see, using two or three shortwave bands instead of one brings the shortwave contribution up so that the OSR dominates the OLR at $p_s = 10$ bar, with a shortwave-dominated regime starting around $T_s = 2400 - 2500$ K. 

Figure \ref{fig:OLR_OSR_tuned} shows band-grey simulations where the averaged shortwave opacity coefficients have been tuned to match the outgoing planetary radiation yielded by \textsc{socrates}. All three subplots then show similar results. At $p_s = 10$ bar, the OSR becomes significant, almost matching the OLR at $T_s = 3000$ K, and at $p_s = 260$ bar, the OSR exceeds the OLR, defining a shortwave-dominated regime starting at $T_s = 2600-2700$ K.

\section{Temperature equilibration in the post-runaway state} \label{sec:discussion}

\subsection{Modeling the shortwave spectrum with band-grey radiation} \label{sec:BG_SW}

The need for tuning the opacity coefficients used in the band-grey model results from the high degree of arbitrariness that averaging over wide spectral regions brings. This is the challenge of handling the shortwave regions when modeled using idealized radiation codes. Averaging over the visible and ultraviolet regions of the spectrum implies ignoring the local variations of the lines, which can lead to an over- or underestimation of the opacity in these regions, depending on the averaging method used. In this work, we tried three cases. The first involves a single shortwave band that includes the visible and ultraviolet regions together. The second has two bands, one that includes the visible range, and the other the ultraviolet range. The third is the same as the second, except that the visible range is split in two bands. 

Each addition of a band multiplies the runtime but increases the accuracy of the opacity in each region. This is illustrated by Tab. \ref{tab:2}, which shows the fact that the tuning coefficients of the shortwave bands become closer and closer to unity as we increase their number. Subdividing the shortwave regions comes down to lessening the need for tuning, meaning that tuning coefficients may get closer to unity, with the limit being using real-gas radiation codes such as \textsc{socrates}, however doing so also increases the runtime significantly.

\subsection{Ocean removal and shift in radiating level} \label{sec:LW_transition}

So long as the outgoing radiation is independent of surface temperature, a planet in a runaway state will continue to get hotter, even after the oceans are completely evaporated. Thus, the factors that eventually allow the outgoing radiation to increase determine the ultimate temperature of the surface, whether a magma ocean forms, and how hot the magma ocean becomes at the surface. 

A critical characteristic of the dew-point adiabat derived from the Clausius-Clapeyron relation, Eq. \eqref{eqn:moist_adiabat}, is that it is independent of the surface temperature. Therefore, as long as the infrared photosphere of the planet is located at a height where the temperature profile is fixed on the saturated moist adiabat of a pure water vapor atmosphere, the outgoing planetary radiation stays constant as the surface warms up during the runaway greenhouse effect. This has been long recognized as the basis of the tropospheric outgoing planetary radiation limit that gives rise to a runaway greenhouse \citep{Nakajima1992,PierrehumbertBook2010,Goldblatt2013}.
As the surface warms further, the moist convective zone shrinks and its lowermost reach gets higher and higher until the atmosphere completely aligns with the dry adiabat. When the saturated zone retreats to higher altitudes, increasing the surface temperature can increase the temperature of the radiating layer, because the radiating layer is connected to the surface temperature by a dry adiabat.
A separate mechanism is that as lower layers get hotter, they can radiate relatively short waves to space through window regions. This mechanism also relies on the distinction between which part of the atmosphere is on the dry adiabat and which part is on the saturated moist adiabat, since if shortwave radiation from a hot part of the atmosphere can radiate through windows, the radiation to space can only increase with surface temperature if the source region of the radiation is in the dry adiabatic part of the atmosphere.
%When the moist convective zone gets thin enough, radiation from the surface can reach space again, which allows the OLR to resume its escape to space. This is illustrated by Figure \ref{fig:ptcff}, representing a set of simulations varying the surface temperature from 600 K to 3000 K. Figure \ref{fig:ptcff}B shows the evolution of the thermal profile of the atmosphere as the surface warms up.

At first, the whole atmospheric column is fixed on the dew point of water, and any layer emitting near-infrared flux upward would do it at a temperature that will not change with surface warming, because of the lack of dependency of the dew-point adiabat on the surface temperature. Then, in the 260 bar case, at around 2000 K, the contribution function seen on Figures \ref{fig:EVSFM10_baseline} to \ref{fig:EVSFM260_tuned} begins to be significant, meaning outside of the moist convective zone, around 0.1 bar and to lower pressures as well as near the surface. In the 10 bar surface pressure case, the surface temperature at which the near-surface layers begin to contribute is higher, at about 2300 K. At this surface pressure, we also see the contribution function getting significant at subsaturated levels at 1500 K, while at $p_s = 260$ bar, this happens at surface temperatures of 2000 K and higher. At higher surface temperatures, the peaks of the contribution function become increasingly located at subsaturated levels, which allows the OPR to regain its dependency on the surface temperature. This effect is visible when plotting the outgoing planetary radiation as a function of the surface temperature, as done in Figure \ref{fig:soc_olr}. The surface contribution $\mathcal{SCF_F}$ derived in section \ref{A1} of the appendix is also plotted. At a surface pressure of 10 bar, it is equal to about $7 \cdot 10^{-6}$, $1.3 \cdot 10^3$, $1.5 \cdot 10^5$, and $1.4 \cdot 10^6$ $\mathrm{W} \: \mathrm{m}^{-2}$ for surface temperatures of 600 K, 1500 K, 2300 K, and 3000 K respectively. Only the SW band including the visible and UV regions is contributing to these values. The bands IR, W1, and W2 are too optically thick to let the surface emission break out to space. Similarly, at the surface pressure of 260 bar, the $\mathcal{SCF_F}$ is respectively $1.9 \cdot 10^{-6}$, $3.4 \cdot 10^2$, $3.8 \cdot 10^4$, and $3.3 \cdot 10^5$ $\mathrm{W} \: \mathrm{m}^{-2}$. At temperatures where the moist convective zone shrinks enough to allow the dominantly contributing layers to become dependent on the surface temperature, beyond 1800 K at $p_\mathrm{s} =$ 260 bar, and beyond 1050 K at $p_\mathrm{s} =$ 10 bar, as we saw with \textsc{socrates}, the OPR is able to increase. Our simulations closely align with the reference curves from \cite{2013ApJ...765..131K}, \cite{Goldblatt2013}, and \cite{2015ApJ...806..216H}. This does not take into account the effect of condensation on radiation, through clouds for instance. Clouds would decrease the level of the radiation limit as well as extend its reach to higher surface temperatures, as \cite{2017JGRE..122.1539M} pointed out, though high altitude clouds would also have a compensating effect in reflecting more incoming stellar radiation back to space.

\subsection{Shift in spectrum to shortwave-dominated} \label{sec:SW_transition}

When the surface warms up to this temperature range and beyond, it starts to radiate increasing amounts of shortwave flux upward, extending to the visible and UV ranges to some extent. In a pure steam environment where water vapor is the only relevant absorber, fluxes at these wavelengths can escape to space through the vapor without being absorbed as much. As this happens, the peak of the Planck function shifts towards the visible range of the spectrum, as seen on Figure \ref{fig:spectral_range_1SWb} B. At about 2000 K the amount of shortwave flux emitted by the surface is on the order of $10 \: \mathrm{W \, m^{-2}}$. Assuming the atmosphere is completely transparent to shortwave radiation, the outgoing shortwave radiation would quickly become a significant addition to the outgoing planetary radiation at the top of the atmosphere, and can bridge the gap to a runaway climate state if conditions are already close enough.

While the processes that makes the OPR increase beyond a certain surface temperature are the same as in Figure \ref{fig:soc_olr}, the OSR yielded by the band-grey model and shown on Figure \ref{fig:OLR_OSR_tuned} increases quite rapidly to dominate the OLR at a surface temperature of about 2730 K at $p_s = 260$ bar case (panel B3). Below this threshold, the outgoing energy of the planet is longwave-dominated; above it, it is shortwave-dominated. At $p_s = 10$ bar, the OSR only reaches about 76\% of the OLR at $T_s = 3000$ K.
Given that the OSR seems to only be dominant at high surface pressures and at temperatures well above 2700 K, these are the conditions that represent the best prospects for observing runaway atmospheres. Lower surface pressures and temperatures would prevent their detection based on the OSR.

This increase has some ramifications for the post-runaway equilibrium state. First, if we assume for instance an incoming stellar radiation of 311 $\mathrm{W} \: \mathrm{m}^{-2}$, the value predicted for Earth in one billion years, it reduces the equilibrium temperature of the post-runaway climate by $\approx$ 18 K in the 260 bar case, which is not significant, and does not change it noticeably at all in the 10 bar case. If only the longwave contribution to the outgoing flux is taken into account, the post-runaway equilibrium temperature would be $T_\mathrm{s} \approx$ 1853 K at 260 bar, where the outgoing longwave radiation intersects the incoming stellar radiation when the Earth is 6.067 Gyr old. Taking into account the shortwave contribution, this shifts the equilibrium to $T_\mathrm{s}$ = 1835 K, where the outgoing planetary radiation intersects the incoming stellar radiation at the same epoch. The instellation received on TRAPPIST-1b is higher, at about 1343 $\mathrm{W} \: \mathrm{m}^{-2}$, assuming the bond albedo is 0. In this case, the equilibrium temperature shift becomes $2451-2364 = 87$ K at 260 bar. Because the outgoing shortwave radiation increases faster than the outgoing longwave radiation, the shift in equilibrium temperature is expected to increase with the instellation. At 10 bar however, the shift is negligible, suggesting that surface pressure also affects the sensitivity of the equilibrium surface temperature to the shortwave contribution.
Rayleigh scattering in the shorter waves would likely reduce the outgoing shortwave radiation, meaning that this cooler equilibrium surface temperature is to be taken as a lower bound. It is worth noting that at the 10 bar surface pressure, the equilibrium surface temperatures are below the solidus temperature of silicates ($T_{eq}$ is found at 1027 K here assuming the same epoch of 6.067 Gyr for the Earth), which means that a solid surface would be able to form and volatile exchange between the atmosphere and the mantle would be reduced.

Both Figs. \ref{fig:soc_olr} and \ref{fig:OLR_OSR_tuned} show how the two different types of runaway greenhouse planets approach the transition. Planets right after formation approach the transition from the right and cool down. This means that for planets within the runaway greenhouse limit \citep{Hamano2013}, the equilibrium temperature during the extended desiccation phase may critically depend on the contribution of the shortwave flux. On the other hand, planets that transition from a possibly habitable state to a runaway phase due to increasing instellation over time approach the transition from the left and are similarly affected by the increase in shortwave flux. For both types of magma ocean planets, primordial and secondary, and within or without the runaway greenhouse limit, the contribution of the shortwave flux and the aforementioned radiation throughout the whole spectrum needs to be considered for future surveys that aim to detect and characterize such magma ocean planets \citep{Lupu2014,Bonati2019}.

The initial volatile inventory of the planet plays a critical role in its evolution. A runaway atmosphere does not start off at the pure steam limit, but approaches it as the water vapor feedback runs its course, however this is assuming that the planet starts off with a large enough initial water reservoir. If not, then the atmosphere may never reach the pure steam limit or even the non-dilute limit, in which case it will not undergo the positive water vapor feedback loop leading up to planet's desiccation. The question of what happens before then, including what might happen if the water inventory is at, say, 1\% of an Earth ocean, warrant further study. With a sufficiently low surface pressure, the equilibrium surface temperature can be low enough that a magma ocean does not form, unless the instellation is very high.

% POSSIBLE ADDITIONS
% - The need for better spectroscopic measurements at high pressures and high temperatures for water vapor, and into the supercritical regime
% - The occurence of the supercritical regime above magma oceans 
% ...

\section{Conclusions} \label{sec:conclusion}

The transition from the runaway to the post-runaway stage is important to understand since it likely marks a dominant evolutionary phase in the history of the rocky exoplanet population, with major implications for their present-day climate, total volatile inventory, and observable features in the atmosphere. Here, we utilized 1-D radiative-convective climate simulations in order to explore this transition phase. Firstly, we explored the change in the main radiating level as a result of an increase in surface temperature during a runaway phase. As the surface warms up, the outgoing radiation begins to become dominated from sub-saturated parts of the atmosphere that follow the dry adiabatic lapse rate, which depends on the surface temperature. Surface warming therefore heats up these atmospheric layers, which increases their contribution to the outgoing radiation. Secondly, we used a computationally-efficient band-grey model with similar settings to evaluate the contribution of the shortwave flux to the outgoing planetary radiation and reproduce  the key features of the full correlated-$k$ model. The flux contribution function is a useful tool that can highlight the parts of the atmosphere that determine the outgoing planetary radiation. For moderate temperatures, the outgoing planetary radiation mostly originates from the upper 10 mb of the atmosphere, indicating the need for adequate vertical resolution in this part of the atmosphere for GCM studies. Surface temperatures above the silicate liquidus, during magma ocean stages in the post-runaway regime, will make the surface and the lower atmosphere radiate in the visible and ultraviolet parts of the spectrum. This may reduce the post-runaway equilibrium temperature of the planet with increasing instellation, and affect the shape of the atmospheric spectrum. Future astronomical observations sensitive to the shortwave regions may probe both the reflection of stellar radiation and innate shortwave radiation from the planet itself.
\acknowledgements
The authors thank J. Manners and D. Amundsen for help with the \textsc{socrates} radiation code, M. Hammond, X. Tan, and V. Parmentier for discussions, E. K. Lee for help with the band-grey radiation model, and J. Kasting and an anonymous reviewer for comments that helped to improve the manuscript. This work was supported by grants from the European Research Council (Advanced grant EXOCONDENSE \#740963 to R.T.P.) and the Simons Foundation (SCOL award \#611576 to T.L.).

\textit{Software:} \textsc{socrates} \citep{edwards1996socrates}, \textsc{numpy} \citep{numpy:2020}, \textsc{scipy} \citep{scipy:2001}, \textsc{pandas} \citep{pandas:2010}, \textsc{matplotlib} \citep{matplotlib:2007}, \textsc{seaborn} \citep{seaborn:2018}.

\vspace{2.3cm}
%\newpage
\bibliography{references}{}
\bibliographystyle{aasjournal}

\appendix
\section{Derivation of the flux contribution function}\label{A1}

We start with the Schwarzschild two-stream radiative equation without scattering,

\begin{equation}\label{eqn:apx1}
\frac{d}{d\tau_\nu}I_\nu(\tau_\nu,\mu,\phi)=-\frac{1}{\mu}\left[I_\nu(\tau_\nu,\mu,\phi)-\pi B(\nu,T(\tau_\nu))\right], 
\end{equation}
with $\mu = \mathrm{cos}(\theta)$ and $\theta$ the zenith angle, and $\phi$ the azimuthal angle. For the rest of the derivation, we will drop the $\nu$ indices for clarity. All quantities depend on frequency. We also use the convention that at the top of the atmosphere $\tau = \tau_\infty$ and at the bottom $\tau = 0$. \cite{Drummond2018} uses the opposite convention. 

The upward flux then satisfies
\begin{equation}\label{eqn:apx2}
\mu\frac{d}{d\tau}I_+(\tau,\mu,\phi)=-I_+(\tau,\mu,\phi) + \pi B(T(\tau)), 
\end{equation}
which, integrated, becomes
\begin{equation}\label{eqn:apx3}
I_+(\tau,\mu,\phi)=I_+(0,\mu,\phi)e^{-\frac{\tau}{\mu}} + \int_{0}^{\tau} \pi B(T(\tau')) e^{-\frac{|\tau-\tau'|}{\mu}}\frac{d\tau'}{\mu}. 
\end{equation}

We then evaluate this expression at the top of the atmosphere.
\begin{equation}\label{eqn:apx4}
I_+(\tau_\infty,\mu,\phi)=I_+(0,\mu,\phi)e^{-\frac{\tau_\infty}{\mu}} + \int_{0}^{\tau_\infty} \pi B(T(\tau')) e^{-\frac{\tau_\infty-\tau'}{\mu}}\frac{d\tau'}{\mu}. 
\end{equation}

Computing the contribution of a given atmospheric layer $\tau_1 \rightarrow \tau_2$ is done by setting the integration bounds to $\tau_1$ and $\tau_2$. The surface contribution is not the contribution of any atmospheric layer, and thus does not come into the expression of the contribution function. It will instead be dealt with separately. \cite{Drummond2018} assume that $B(T(\tau'))$ is constant over the layer $\tau_1 \rightarrow \tau_2$ of width $d\tau'$. We keep this assumption and define $\tau_{12}$ any optical depth in the range $[\tau_1,\tau_2]$.

The contribution of a particular layer $\tau_1 \rightarrow \tau_2$ is then

\begin{equation}\label{eqn:apx5}
I_+(\tau_\infty,\mu,\phi)=\pi B(T(\tau_{12})) \int_{\tau_1}^{\tau_2} e^{-\frac{\tau_\infty-\tau'}{\mu}}\frac{d\tau'}{\mu}, 
\end{equation}

with

\begin{equation}\label{eqn:apx6}
\int_{\tau_1}^{\tau_2} e^{-\frac{\tau_\infty-\tau'}{\mu}}\frac{d\tau'}{\mu} = \frac{\mu}{\mu}e^{-\frac{\tau_\infty}{\mu}}\left(e^{\frac{\tau_2}{\mu}} - e^{\frac{\tau_1}{\mu}}\right)=\left(e^{-\frac{\tau_\infty-\tau_2}{\mu}} - e^{-\frac{\tau_\infty-\tau_1}{\mu}}\right),
\end{equation}

The intensity contribution function $\mathcal{CF}_I$ is then

\begin{equation}\label{eqn:apx8}
\mathcal{CF}_I=I_+(\tau_\infty,\mu,\phi)=\pi B(T(\tau_{12})) \left(e^{-\frac{\tau_\infty-\tau_2}{\mu}} - e^{-\frac{\tau_\infty-\tau_1}{\mu}}\right). 
\end{equation}

The flux contribution function $\mathcal{CF}_F$ is

\begin{equation}\label{eqn:apx9}
\mathcal{CF}_F=\int_{0}^{2\pi}d\phi \int_{0}^{1} \mu\mathcal{CF}_I d\mu,
\end{equation}

Replacing, we get

\begin{equation}\label{eqn:apx10}
\mathcal{CF}_F=2\pi \pi B(T(\tau_{12})) \int_{0}^{1} \left(\mu e^{-\frac{\tau_\infty-\tau_2}{\mu}} - \mu e^{-\frac{\tau_\infty-\tau_1}{\mu}} \right)d\mu.
\end{equation}

%The surface contribution function is simply

%
%\begin{equation}\label{eqn:apx11}
%\mathcal{SCF}_F=2\pi\int_{0}^{1}I_+(0,\mu) \mu e^{-\frac{\tau_\infty}{\mu}}\delta(\tau)d\mu.
%\end{equation}
%end{linenomath*}

This is solved with the diffusivity approximation (\cite{2002rtao.book.....T}, section 11.2.5) providing

\begin{equation}\label{eqn:apx12}
\int_{0}^{1} \mu e^{-\frac{\tau}{\mu}}d\mu=\frac{e^{-D\tau}}{D},
\end{equation}

with $D \approx 1.66$ the diffusivity factor.

The $\mathcal{CF}_F$ becomes

\begin{equation}\label{eqn:apx13}
\mathcal{CF}_F = 2\pi \pi B(T(\tau_{12}))\frac{e^{-D\frac{\tau_\infty - \tau_2}{\mu}}-e^{-D\frac{\tau_\infty - \tau_1}{\mu}}}{D}.
\end{equation}

Assuming that $I_+(0)$ is independent of $\mu$, we get

\begin{equation}\label{eqn:apx14}
%\begin{split}
\mathcal{CF}_F = 2\pi \pi B(T(\tau_{12}))\frac{e^{-D(\tau_\infty - \tau_2)}-e^{-D(\tau_\infty - \tau_1)}}{D}
%\end{split}
\end{equation}

The surface contribution function is simply the upward flux radiated by the surface attenuated by the entire atmospheric column.

\begin{equation}\label{eqn:apx11}
\mathcal{SCF}_F=I_+(0) e^{-\tau_\infty}.
\end{equation}

The column-integrated $\mathcal{CF}_F$ should yield the outgoing radiation. Let's integrate it from 0 to $\tau_\infty$. First, we express the exponentials in the $\mathcal{CF}_F$ expression of Equation \eqref{eqn:apx14} as a differential, as

\begin{equation}\label{eqn:apx15}
\mathcal{CF}_F = \frac{2\pi}{D} \pi B(T(\tau_{12})) de^{-D(\tau_\infty - \tau)}.
\end{equation}

We can solve it by integrating by parts:

\begin{equation}\label{eqn:apx16}
\begin{split}
\int_{0}^{\tau_\infty} \mathcal{CF}_F d\tau' &= \frac{2\pi}{D} \left[\pi \left(B(T(\tau_\infty))e^{-D(\tau_\infty - \tau_\infty)}-B(T(0))e^{-D(\tau_\infty - 0)}\right) - \int_{0}^{\tau_\infty}\pi dB(\tau)e^{-D(\tau_\infty - \tau)} \right]\\
&= \frac{2\pi}{D} \left[\pi \left(B(T(\tau_\infty))-B(T(0))e^{-D\tau_\infty}\right) - \int_{0}^{\tau_\infty}\pi dB(\tau)e^{-D(\tau_\infty - \tau)} \right]
\end{split}
\end{equation}

The outgoing radiation yielded by the Schwarzschild radiative equation is

\begin{equation}\label{eqn:apx17}
I_+(\tau_\infty) = I_+(0)e^{-\tau_\infty} + \pi \left(B(T(\tau_\infty)) - B(\tau_0)e^{-D\tau_\infty}\right) - \int_{0}^{\tau_\infty} \pi dB(T(\tau)) e^{-D(\tau_\infty - \tau)},
\end{equation}

which is equal to the integrated $\mathcal{CF}_F + \mathcal{SCF}_F$ except for the $\frac{2\pi}{D}$ factor due to the angle integration and the diffusivity approximation. 

\end{document}